\documentstyle[rmaaconf,psfig]{article}

\begin{document}

\title{MASS SEGREGATION IN VERY YOUNG OPEN CLUSTERS}

\author{Didier Raboud}

\affil{Geneva Observatory, CH-1290 Sauverny, Switzerland}

\begin{resumen}
Escribir aqu\'{\i} el  resumen en espa\~nol (generalmente es 10\% 
m\'as largo que en ingl\'es).

...

...

...

...
\end{resumen} 

\begin{abstract}
  The study of the very young open cluster NGC 6231 clearly shows the presence of a
  mass segregation for the most massive stars. These observations, combined with those
  concerning other young objects and very recent numerical simulations, strongly
  support the hypothesis of an initial origin for the mass segregation of the most
  massive stars. These results led to the conclusion that massive stars form
  near the center of clusters. They are strong constraints for scenarii of star and
  stellar cluster formation.
\end{abstract}

\bigskip

\noindent

\section{Introduction - Data - Results}

The aim of the present work is to give observational constraints to better
understand the dynamical evolution of open clusters (OCs). For an analysis of this
evolution the observation and study of clusters of different ages is required.
In order to fix the "initial conditions" we chose to extensively observe NGC 6231,
which is a rich and very young OC: it has an age of 3-4 Myr and contains more than
one hundred O and B stars (Raboud 1996, Raboud et al. 1997). 
The data collected allowed us to complete the analysis of this cluster structure
(Raboud 1997, Raboud \& Mermilliod 1998, RM98).

\section{Discussion}

\begin{figure}[thb]
\centerline{\psfig{figure=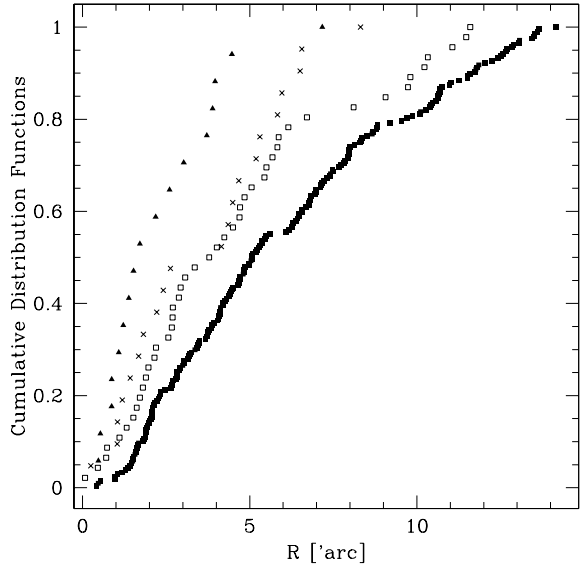,height=80mm,width=80mm}}
\caption[]{Cumulative distributions for stars of different mass intervals, as a
function of the distance to the center of the cluster NGC 6231. The most massive
stars are more concentrated toward the cluster center. Triangles: $M >$ 16
M$_\odot$, crosses: 10 $< M <$ 16 M$_\odot$, open squares: 6 $< M <$ 10 M$_\odot$,
filled squares: $M <$ 6 M$_\odot$.}
\label{fig:ms6231}
\end{figure}

The main result of our study is the observation of a mass segregation in 
NGC 6231 which can be clearly seen in Fig. \ref{fig:ms6231}. NGC 6231 being a 
very young OC, it was important to first solve the problem of the origin of
the mass segregation: is it \textit{initial} (i.e. the imprint of the stellar
formation processes) or is it the consequence of the dynamical evolution of the
cluster ?

As a first step towards the understanding of this problem we computed the mean
relaxation time ($t_{r}$) of the cluster. Using the standard equations from Chandrasekhar
(1942) and Spitzer \& Hart (1971) we derived a value of $t_{r} \approx 10^{7}$ yr.
This result is a lower limit because we observed only the brightest stars of the
cluster and therefore underestimated the total number of stars and 
the characteristic radius of the cluster while overestimating its mean stellar mass.
As the age of the cluster is nearly one order of magnitude smaller than its 
$t_{r}$, we conclude that the observed mass segregation in NGC 6231 is an
imprint of the stellar formation processes.

Nevertheless, the preceding conclusion is dependent on the physical validity of
the \textit{mean} $t_{r}$. This time refers to stars of average mass. As
real clusters present a wide mass spectrum, this implies that the systems
evolve on a timescale shorter than that estimated by this mean $t_{r}$. 
Furthermore, $t_{r}$ depends upon the location in the cluster: it
significantly increases from the center to the outer regions.
Moreover, $N$-body calculations that treat close gravitational encounters and
binary formation predict more rapid dynamical evolutions, typically one order
of magnitude, than that indicated by the mean $t_{r}$. Therefore $t_{r}$ 
has to be an upper limit and the conclusion stated above that the observed mass
segregation in NGC 6231 is not a consequence of the dynamical evolution of the cluster
is strongly weakened.

However, numerous other pieces of observational evidence for initial mass
segregation in OCs exist in the Galaxy and the LMC. But, as these ``proofs''
are mainly based on the comparison between the age of the cluster and its mean
$t_{r}$, these studies suffer drawbacks similar to those described above.

Then, what would be the solutions to unambigously reveal an \textit{initial}
mass segregation ? There are two possibilities: observational \textbf{(a)}
and numerical \textbf{(b)}.

\textbf{(a)} From the observational point of view we have to observe not only
very young OCs but also \textit{extremely} young OCs, i.e. clusters with ages
of the order of their crossing time or below. In such clusters relaxation
processes have no meaning and the observed locations of the stars are
close to their birthplace. Examples are, among others,  NGC 2024 and NGC 
2071 (Lada \& Lada 1991). These clusters are still embedded in their
parental cloud and already present a mass segregation.

\textbf{(b)} From the numerical point of view we could simulate a cluster and
test if an observed mass segregation could possibly be explained
by relaxation processes or if we need the ''help'' of an initial segregation.
Such a modelling had been done by Bonnell \& Davies (1997) for the
Orion Nebula Cluster (ONC). The authors show that the position of massive stars in
the center of rich young clusters cannot be due to dynamical mass segregation.
In particular, they claim that for producing a Trapezium-like system within
just a few crossing times, the massive stars most likely formed within the
inner 10\% of the cluster.

These last considerations tend to favor the reality of, at least partially,
initial mass segregation in very young OCs. This result implies that the
IMF is, locally, not unique. We observe its variation: it is flatter in the
central part of the cluster and steeper in the outer part.

Furthermore, a closer inspection of Fig. \ref{fig:ms6231} reveals that only
the most massive stars ($M >$ 16 M$_\odot$) are clearly concentrated toward the
cluster center. The stars belonging to the two intermediate mass intervals
(6 $< M <$ 16 M$_\odot$) are spatially  well mixed.
Similar results are obtained for a cluster embedded in the MonR2 cloud (Carpenter
et al. 1997). Moreover, in the case of the ONC, Fig. 6 from Hillenbrand (1997)
shows very different spatial distributions for stars more massive or less
massive than 5 M$_\odot$. For masses smaller than 5 M$_\odot$ the distributions are
rather similar.

\section{Conclusion: Double origin for the mass segregation ?}

The above results allow us to propose a qualitative scenario for the evolution of
mass segregation with age in OCs (RM98):

\bigskip

\textbf{(I)} The most massive stars form in the center of clusters.

Several hypotheses could be made to explain this phenomenon:
dynamical friction between protostellar clouds and inter-protostellar medium
(Larson 1991, Gorti \& Bhatt 1995, 1996); collision and coalescence of
protostellar clouds (Murray \& Lin 1996); the accretion of matter during stellar
formation phases. This accretion could be faster in regions of higher temperature and
turbulence (Maeder 1997), i.e. in the center of protocluster clouds,
thus leading to the formation of more massive stars in these regions.
This last hypothesis implies that the IMF is dependent on the local physical conditions.

In the context of massive star formation in the center of clusters, it is worth
noting that we observe numerous examples of multiple systems of O-stars in the
center of very young OCs. In the case of NGC 6231, 8 stars among the 10 brightest
are spectroscopic binaries with periods shorter than 6 days. Moreover, we observe
trapezium systems of O-stars in the ONC, NGC 6823 and Tr 37. Four-component and
triple systems have also been found in NGC 2362 and Collinder 228.

\textbf{(II)} In less than $10^{7}$ yr these spatially concentrated massive stars
will disappear due to stellar evolution. As they represent
a non-negligible percentage of the total mass of the cluster (between $\sim$10
and 30 \% in the case of NGC 6231), the disappearance of these massive
stars could lead to a violent relaxation phase. If a mass segregation was
previously established in the cluster it could be more or less erased during
this phase, depending on the importance of the initial population of massive
stars.

We are then {\it possibly} left with a cluster presenting {\it no} mass segregation
at all.
NGC 6531 (Forbes 1996) provides an example of such a cluster: it is
8 $\times$ 10$^{6}$ yr old and does not contain any stars with masses greater than
20 M$_\odot$, which make up the concentrated population in NGC 6231. Forbes shows
convincingly that NGC 6531 does not exhibit any mass segregation, 
and he explains his observation by the young age of the cluster. According
to him, NGC 6531 is too young for dynamical evolution to have left any significant 
impression. But this hypothesis was based on an estimation of $t_{r}$ and suffers the
drawbacks described in the Discussion.

Another interesting point related to the disappearance of the massive stars
is the stability of the cluster. It is possible that a bound cluster
becomes unbound after this violent phase. Numerical simulations by Terlevich
(1987) show that clusters with flat initial mass functions have to be rich enough to
survive the initial violent period of mass loss.

\textbf{(III)} The last point of our scenario is that all  mass segregation
observed in older clusters is \textit{merely} the consequence of the cluster's
\textit{dynamical evolution}.

\bigskip

To better quantify this hypothesis of a possible double origin (initial
and dynamical) of the mass segregation we need to analyse the structure of OCs old
enough (around 10$^{7}$ yr) to have lost their most massive stars. 
Thus, one consequence of our hypothesis is that
some of these clusters, those which initially contained an important population  
of massive stars, should not present any mass segregation.

\end{document}